\documentstyle[aps,tighten,epsfig]{revtex}
\begin{document}
\draft
\title{Andreev reflection in superconducting QCD}
\author{ Mariusz Sadzikowski$^{1,2}$ and Motoi Tachibana$^{3}$ }
\address{ 1) Center of Theoretical Physics, Massachusetts Institute of Technology,
             Cambridge, MA 02139, USA\\
          2) Institute of Nuclear Physics, Radzikowskiego 152,
             31-342 Krak\'ow, Poland\\
          3) Theoretical Physics Laboratory, RIKEN, 2-1 Hirosawa,
             Wako, Saitama 351-0198, Japan}
\maketitle
\begin{center}
MIT-CTP-3289
\end{center}

\begin{abstract}
In this paper we discuss the phenomenon of the Andreev reflection 
of quarks at the interface between the 2SC and the Color-Flavor-Locked 
(CFL) superconductors appeared in QCD at asymptotically high densities.
We also give the general introduction to the Andreev reflection in the
condensed matter systems as well as the review of this subject in high
density QCD.
\end{abstract}

\section{Introduction}

Quantum Chromodynamics is a non-abelian quantum field theory describing
interactions between quarks and gluons. It describes the effect of
confinement at low energy \cite{lattice} and asymptotic freedom at high energy \cite{gw_p}
compared to the QCD scale $\Lambda_{QCD}\sim 200$ MeV. The only known
systematic approach to low energy QCD are lattice calculations. The 
physics of the high energy QCD can be controled by the perturbation
analysis in strong coupling constant. There are three different types of phenomena
of strong interactions which one can hope to describe in the perturbative regime:
high energy scattering processes, high temperature and high baryon density systems.
The energy scale for these phonomena are set by: the value of the four-momentum transfer 
$\sqrt{Q^2}$ exchanged in the process, the value of temperature $T$ and the value 
of the Fermi energy $E_F$ (or quark chemical potential $\mu $) respectively.
The high energy scattering processes were already well tested in deep inelastic
scattering of leptons on protons or electron-positon annihilation
(DESY, CERN). The samples of the matter at high temperature ($T\sim 200-300$ MeV) 
and low baryon density one can hope to achieve in high energy ions colliders (RHIC). 
However lattice calculations show that the expected temperatures are too low
by the factor of 3-4 for the application of the perturbative QCD.
Then one has to rely on the non-perturbative analysis: lattice
calculations, universal features of phase transitions or models.
Finally the high baryon density quark matter at low temperatures is expected
to exist in the cores of the compact stars \cite{glend}. Unfortunately, here also the densities
are not high enough for direct perturbative calculations (by many orders
of magnitude). Then the only possible approches are based on universal features of
the phase transitions and models. The lattice approach fails at non-zero density
physics because of the sign problem. 

In this paper we focus on the high baryon density phases at $T=0$. 
Since long ago one expected supeconducting phenomena
in this kinematical region \cite{bailin}. However only recently
there have been a new interest in the subject because new features
were found in the theory \cite{arw1_rssv,arw2,son}. 
If the energy scale is set by the quark chemical potential $\mu$
one can expect that at high enough densities there are a gas
of free quarks and gluons. However the interaction between quarks
mediated by the one-gluon exchange is attractive in the
color $\bar{\mbox{\bf 3}}$ channel. This leads to the well known
phenomenon of the Cooper instability of the Fermi sea and finally
results in the creation of a new vacuum - the condensate of Cooper pairs.
This is the version of the BCS theory of superconductivity \cite{bcs} applied
to the interacting quark matter. The picture is generally correct,
however somehow oversimplified. There are subtle but important differences
between BCS theory and superconductivity in QCD which follow
from the non-abelian character of the interation in quark-quark
scattering (eg. \cite{son}). The review of the whole subject can be found in \cite{rw}.
Despite of some differences the high density QCD shares a lot of features with
condensed matter systems. Indeed at high density the relativistic
quantum field theory effects are suppressed and one can apply
just non-relativistic field theory or relativistic quantm mechanics 
to study interesting phenomena. 

Our subject of interest are scattering processes 
at the interfaces between different types of QCD phases. These are the analogs of
the Andreev reflection \cite{andreev} in condensed matter systems.
This reflection apperas at the junction between conductors
and superconductors. In the case of dense QCD the role of conductor 
is played by the free Fermi gas of quarks and
superconductor is 2SC or CFL phase of QCD. 
In this paper we consider the interface between 2SC and CFL phase.
We shall show that this interface is the most general in a sense
that it already contains the other cases: free quarks/2SC \cite{sad} 
and more: 2SC/2SC' and one only peculiar to 2SC/CFL.
The case of free quarks/CFL interface \cite{sad_tac} has to be considered independently.

The interest in the dynamics of the matter flow through
the phase boundaries is of importance because such interfaces can be present in
the protoneutron stars.
Indeed in the final stage of neutron star cooling one expects that the
star can go through the all possible phase transitions
\cite{carter_reddy} including 2SC/CFL case. This last
transition is expected to be the first order \cite{abr} thus
one expects the existence of the mixed phase: CFL-bubbles in
the 2SC medium as well as 2SC-bubbles in the CFL medium. 
In such the conditions the Andreev reflection
is frequent phenomena and should be taken into account
in the calculations of detail processes of the protoneutron
star evolution.

In the sections 2 and 3 we provide the general introduction
to superconductivity and Andreev reflection in
condensed matter systems. Section four contains
short description of 2SC and CFL phases,
whereas in Section five we describe the Andreev
reflection in QCD. There are also three Appendices
which gives more detailed calculations.

\section{Conductors and superconductors}

Let us start with the simple effective hamiltonian describing superconducting
state of the system at zero temperature:
\begin{equation}
\label{hamiltonian}
H = \int d^3x\left[\sum_{\alpha =\uparrow ,\downarrow }\psi^{\dagger }_\alpha (t,\vec{r})
\left( -\frac{\nabla^2}{2 m} - E_F\right)\psi_\alpha (t,\vec{r}) +
\Delta (\vec{r})\psi^{\dagger}_\uparrow (t,\vec{r})\psi^{\dagger}_\downarrow (t,\vec{r})
+ \Delta^\ast (\vec{r})\psi_\downarrow (t,\vec{r})\psi_\uparrow (t,\vec{r})\right] .
\end{equation}
The fermi field operator $\psi_\alpha (t,\vec{r})$ ($\psi^{\dagger }_\alpha (t,\vec{r})$)
describes the annihilation (creation) of the particle of mass $m$ and spin $\alpha $ at
the point $\vec{r}$ and the time $t$. The particles also carry charge however
in our simple example we do not consider the electromagnetic fields thus this
feature is unimportant for our purpose. The quantity $E_F$ defines the Fermi surface
below which all energy levels are occupied in accordance with Pauli principle.
The so called pair potential $\Delta (\vec{r})$ varies in space and describes the
superconducting gap in excitation spectrum of the system - the essence of
superconductivity. The last 2 terms in the above effective hamiltonian are
a mean-field approximation of attractive four-fermi point interaction. 
The pair potential $\Delta $ is related to the anomalous particle-particle correlator: 
\begin{equation}
\Delta_{\alpha\beta }(t,\vec{r}) = \langle\psi_\alpha (t,\vec{r})\psi_\beta (t,\vec{r})\rangle = 
\Delta (t,\vec{r})\epsilon_{\alpha\beta } .
\end{equation}
One can treat this hamiltonian as the description of the interation between
particles $\psi $ and given classical pair potential $\Delta (\vec{r})$. 
We assumed that this interaction is independent of the spin. In that sense
the spin is not dynamical degree of freedom but only a quantum number that 
distinguishes between 2 types of particles.

Let us mention that the hamiltonian
(\ref{hamiltonian}) does not conserve particle number\footnote{In this
situation charge is also not conserved.}. This results from the
fact that the pair potential describes the condensate of Cooper pairs which is
a source of particles in our approximation. Let us also notice that the
energy is conserved however not necessarily momentum. The momentum conservation
is related to the space dependence of the gap parameter $\Delta (\vec{r})$.

The equations of motion that follow from the hamiltonian (\ref{hamiltonian}) and
standard anticomutation relations for fermi fields take the form of so called
Bogoliubov - de-Gennes equations \cite{bogoliubov}:
\begin{equation}
\label{eom}
i\partial_t\left(\begin{array}{c}
\psi_\uparrow (t,\vec{r})\\
\psi^{\dagger }_\downarrow (t,\vec{r})
\end{array}\right) = 
\left(\begin{array}{cc}
-\frac{\nabla^2}{2m}-E_F & \Delta(\vec{r}) \\
\Delta^\ast (\vec{r}) & \frac{\nabla^2}{2m}+E_F 
\end{array}\right)
\left(\begin{array}{c}
\psi_\uparrow (t,\vec{r})\\
\psi^{\dagger }_\downarrow (t,\vec{r})
\end{array}\right) .
\end{equation}
These operator equations are linear which means that we are free to treat them
as the usual wavefunction equations (hermitian conjugate changing to complex
conjugate). The physical interpretation of the wavefunctions comes
from the following reasoning. In the case of $\Delta = 0$ the equations (\ref{eom})
decouple from each other:
\begin{eqnarray}
i\dot{\psi}_\uparrow (t,\vec{r}) = \left( -\frac{\nabla^2}{2m}-E_F\right)\psi_\uparrow (t,\vec{r})\\\nonumber
i\dot{\psi}^{\ast }_\downarrow (t,\vec{r})= 
\left(\frac{\nabla^2}{2m}+E_F\right)\psi^{\ast }_\downarrow (t,\vec{r}) .
\end{eqnarray}
The plane wave solutions:
\begin{equation}
\phi\equiv\left(\begin{array}{c}
\psi_\uparrow (t,\vec{r}) \\
\psi^{\ast }_\downarrow (t,\vec{r}) 
\end{array}\right) =
\left(\begin{array}{c}
f \exp (-iEt + i\vec{q}\vec{r})\\
g \exp (-iEt + i\vec{q}\vec{r})
\end{array}\right)
\end{equation}
lead to the dispersion relations:
\begin{equation}
\begin{array}{lr}
E = \epsilon_q & \mbox{for particles}\\
E = - \epsilon_q & \mbox{for holes} ,
\end{array}
\end{equation}
where $\epsilon_q = \vec{q}^{\, 2}/2m - E_F$ is a measure of the
kinetic energy with respect to Fermi level and $E$ describes the 
energy excitation above (below) Fermi energy
for particles (holes)\footnote{The apparent correlation between
spin and particle-hole interpretation is accidental. We can
have the same set of equations with exchanged spin indices.}. 
Let us mention the interesting detail that for the holes the group velocity is opposite
to momentum $\vec{v} = - \vec{q}/m$ thus the wavefunction
$\psi^{\ast }_\uparrow (t,\vec{r})$ describes the hole which is moving
from right to left for $\vec{r}=(0,0,z)$\footnote{One can also say that
the hole has a negative mass.}. 
The above physical picture can be treated as a simple description
of the conductors. The energy spectrum contains particles
and holes in the vicinity of Fermi surface which can be excited easily 
for the negligible cost of energy.

Let us now consider equations (\ref{eom}) with $\Delta (\vec{r}) = const.$,
which gives us the model of superconductor.
The plane-waves are still the solutions of (\ref{eom}):
\begin{equation}
\phi = \left(\begin{array}{c}
f_q \exp (-iEt + i\vec{q}\vec{r})\\
g_q \exp (-iEt + i\vec{q}\vec{r})
\end{array}\right) ,
\end{equation}
where momentum dependent constants satisfy the equations:
\begin{equation}
\label{eom_quasi}
\left(\begin{array}{cc}
-E+\epsilon_q & \Delta \\
\Delta^\ast & -E-\epsilon_q
\end{array}\right)
\left(\begin{array}{c}
f_q\\
g_q
\end{array}\right) = 0 .
\end{equation}
The selfconsistency condition for the set of homogenous equations gives
the dispersion relation:

\begin{equation}
E^2 = \epsilon_{q}^{\, 2}+|\Delta |^2 .
\end{equation}
This is famous gapped energy spectrum of the quasiparticles
in the superconductor.
The wavefunction describing quasiparticles are given by the solutions
of (\ref{eom_quasi}):
\begin{equation}
\label{quasi}
\phi =
D\left(\begin{array}{c}
\sqrt{\frac{1}{2}(1+\xi /E)}\exp (i\delta /2)\\
\sqrt{\frac{1}{2}(1-\xi /E)}\exp (-i\delta /2)
\end{array}\right)
\exp (i\vec{q}_+\vec{r}-iEt) +
F\left(\begin{array}{c}
\sqrt{\frac{1}{2}(1-\xi /E)}\exp (i\delta /2)\\
\sqrt{\frac{1}{2}(1+\xi /E)}\exp (-i\delta /2)
\end{array}\right)
\exp (i\vec{q}_-\vec{r}-iEt) ,
\end{equation}
where $\xi = \sqrt{E^2-|\Delta |^2}$, $\delta $ is a phase of the gap parameter,
$D,F$ some constants and momenta of the excitations are given by the relations:
\begin{equation}
\frac{\vec{q}_{\pm}^2}{2m} = E_F\pm\xi .
\end{equation}
The first (second) wave-function of (\ref{quasi}) is particle-like
(hole-like) excitation because for $E>>|\Delta |$ it reduces to the particle (hole)
wavefunction. Let us notice that for the excitation with energy
$E<|\Delta |$ the momenta are complex and thus quasiparticles
do not propagate in the medium in such a case. 

\section{Andreev reflection in condensed matter systems}

Let us consider the conductor - superconductor juction (NS). This
can be modeled by the equations (\ref{eom}) when the pair
potential takes appropriate shape. In the simplest case one
can consider the plane junction perpendicular
to the $z$ - axis located in the point $z=0$. Then the gap
parameter is only a function of $z$ coordinate $\Delta (\vec{r}) = \Delta (z)$. 
The exact form of the pair potential depends on the details of the junction
however some general features can be already obtained in the simple model
of the step junction $\Delta (z) = \Delta\Theta (z)$ which describes
a conductor for $z<0$ and superconductor with a constant gap $\Delta $ for $z>0$.
The task is to solve the stationary problem of coupled Schrodinger equations
(\ref{eom}) with boundary conditions:
\begin{enumerate}
\item For $z\rightarrow -\infty $ excitations are particles or holes (conductor)
\item For $z\rightarrow +\infty $ excitations are quasiparticles (superconductor)
\item For $z=0$ wavefunctions and their first spatial derivatives are continuous.
\end{enumerate}
For definitness let us consider that the particle of given energy $E$
incoming from the left hits the junction at $z=0$. After the collision
the particle and/or hole can be reflected back and the qusiparticles
can be excited in superconductor. Thus for the conductor side ($z<0$):
\begin{equation}
\phi_< = \left(\begin{array}{c}
\exp (ikz)+B\exp (-ikz)\\
C\exp (ipz)
\end{array}\right)
\end{equation}
where $k = \sqrt{2m(E_F+E)},\,p= \sqrt{2m(E_F-E)}$ and the constants $B$ and $C$
are related to the probability of the reflection for the particle and hole, respectively.
For the superconductor side ($z>0$) the wave-functions are given by
formula (\ref{quasi}) with $\vec{r} = (0,0,z)$. 
The momenta are given by: $q_{\pm}=\pm w+i m|\xi |/w$
where $w=\sqrt{m(E_F+\sqrt{E_{F}^2-\xi^2})}$ for $E<\Delta $ and 
$q_\pm = \pm\sqrt{2m(E_F\pm\xi)}$ for $E>\Delta $.
The constants $D$ and $F$ describe the probability of the
particle-like and hole-like quasiparticles excitations. The continuity
conditions at $z=0$ lead to the 4 equations for 4 constants $B,C,D$ and $F$.
The solutions of these equations are:
\begin{eqnarray}
B=O(\frac{1}{E_F})\\\nonumber
C=\sqrt{\frac{E-\xi }{E+\xi }} + O(\frac{1}{E_F})\\\nonumber
D=\sqrt{\frac{2E}{E+\xi }} + O(\frac{1}{E_F})\\\nonumber
F=O(\frac{1}{E_F})
\end{eqnarray}
Thus one can conclude that the dominant contribution comes from the hole reflection
in conductor and particle-like excitation in superconductor. The other contributions
are suppresed by the powers of Fermi energy. For the interpretation of our
result let us consider the transition coefficients. From the equations (\ref{eom})
follows the conservation of probability current:
\begin{equation}
\frac{\partial }{\partial t}(|\psi_\uparrow |^2+|\psi_\downarrow |^2) +
\vec{\nabla }\cdot\frac{1}{2mi}\left(\psi^{\ast }_\uparrow\vec{\nabla }\psi_\uparrow
-\psi_\uparrow\vec{\nabla }\psi^{\ast }_\uparrow + \psi_\downarrow\vec{\nabla }\psi^{\ast }_\downarrow -
\psi^{\ast }_\downarrow\vec{\nabla }\psi_\downarrow\right) = 0.
\end{equation}
This probability current through the NS junction has a form:
\begin{equation}
\label{nr_current}
j_z =\left\{\begin{array}{lcr}
0 &\;\;& \mbox{for}\,\, E < |\Delta | \\
\frac{2\xi }{E+\xi }v_F + O(\frac{1}{E_F}) &\;\;& \mbox{for}\,\, E > |\Delta |
\end{array}\right.
\end{equation}
where $v_F = \sqrt{2E_F/m}$  is Fermi velocity. The result of the vanishing 
current for energies below the gap is exact. The transition and reflection 
coefficients read:
\begin{eqnarray}
\label{nr_rt}
R_{hole}=1,\;\;\; T_{quasi} = 0\;\;\;\;\mbox{for}\,\, E< |\Delta | \\\nonumber
R_{hole} = \frac{E-\xi }{E+\xi },\;\;\; T_{quasi} = \frac{2\xi }{E+\xi }\;\;\;\;
\mbox{for}\,\, E > |\Delta |
\end{eqnarray}
Thus for the energies of the incoming particle below the gap only the hole is reflected.
This is easy to understand because there is no possibility to transport energy
into the superconductor below the gap. Let us notice that in the Andreev reflection
the charge is not conserved. This follows from the fact that our hamiltonian
does not conserve particle number. Obviously at the microscopic level the charge is conserved.
The interpretation of the process is simple: the incoming particle creates the Cooper pair
with another particle from the Fermi sea and "dilutes" in the condensate. The lack
of the second particle is visible as a hole. Additionally the Andreev reflection process conserves
the energy and momentum (up to $1/E_F$ corrections). It is also called retro-reflection
because the outgoing hole after reflection is a time reversal picture of the incoming 
particle before the reflection. This is
approximate picture exact only in the limit of the vanishing particle energy $E=0$.
Although the probability current vanishes at the NS junction the charge current
is excessed. Indeed the charge current at the conductor side is double
because the particle and hole currents add to each other. The charge current
is conserved at the interface thus at the superconducting side we also have
to encounter doubled charge current. This current is carried by the condensate
(supercurrent) itself because the quasiparticles penetrate only on the
depth of the order $v_F/|\xi |$.

For the energy of the incoming particle above the gap the coefficients 
$R_{hole}$ and $T_{quasi} $
give us the probabilities of the reflection of the hole or the transmission of
the quasiparticle into superconductor. For $E>>|\Delta |$ the coefficients 
$R_{hole}\rightarrow 0, T_{quasi}\rightarrow 1$ as it should be. 

These predictions were checked experimentally many times, e.g., by the direct measurement
of the reflected holes \cite{benistant}. The Andreev reflection can influence transport
processes at the NS junctions which were already shown by Andreev \cite{andreev}.
Other effects like excess current, charge imbalance and supercurrent conversion
were also considered (e.g., \cite{blonder}).

\section{Superconductivity in QCD}

The perturbative approach to high density QCD is well established 
at much higher densities than one can expect in the cores of the compact stars. 
Thus the use of models are inevitable in the description of interesting physics. 
This is the approach we chose in this paper. 
The one gluon exchange suggests that the diquark condensate is created in
color and flavor anti-symmetric scalar channel. Similar feature
one can also find using instanton-based models.
The effective hamiltonian describing quark interaction with the pseudoscalar 
condensate at the high baryon density can be written in the general form
(for the review see \cite{rw}):
\begin{equation}
\label{hamiltonian_qcd}
H = \int d^3x \left[\sum_{a,i}\psi^{i\, \dagger }_{a}
(-i\vec{\alpha }\cdot\vec{\nabla } - \mu )\psi^{i}_{a}
 +\sum_{a,b,i,j}\Delta^{ij\,\ast}_{ab} (\psi^{i\, T}_{a}C\gamma_5\psi^{j}_b) + \mbox{h.c}\right]
\end{equation}
where $\psi^{i}_a $ are Dirac bispinors, $a,b$ color indices, $i,j$ flavor indices,
$C$ the charge conjugation matrix and $\mu $ is the quark chemical potential.
At zero temperature the condensate $\Delta^{ij}_{ab}$ is a
vacuum expectation value of the 2-point field correlator:
\begin{equation}
\label{gap}
\langle\psi^{i\, T}_aC\gamma_5\psi^{j}_b\rangle = \Delta^{ij}_{ab}
\end{equation}
The formula (\ref{hamiltonian_qcd}) is a relativistic generalization of
the hamiltonian (\ref{hamiltonian}) with the more complicated gap structure.

In the case of two flavors, the gap matrix (\ref{gap}) becomes:
\begin{equation}
\Delta^{ij}_{ab} = \tilde{\Delta }\epsilon^{ij}\epsilon_{ab3}
\end{equation}
where the third direction in color space was chosen
arbitrarily. This combination breaks gauge symmetry
$SU(3)_c\rightarrow SU(2)_c$ whereas chiral symmetry 
$SU(2)_L\times SU(2)_R$ remains untouched. This is the 2SC phase. 
The Cooper pairs are created between the quarks of different
two colors and flavors and third color quarks are unpaired. The lowest energy excitations
are of course unpaird quarks. The quasiparticles are separeted
by the gap $\tilde{\Delta }$ from the vacuum. The value of the
gap depends on the model and is usualy in the range $50-150$ MeV.

For the case of three flavors the gap parameter takes the form\footnote{There is
also a small admixture of the condensate in color symmetric {\bf 6} channel but we neglect 
it in our considerations.}:
\begin{equation}
\Delta^{ij}_{ab} = \Delta\epsilon^{ijk}\epsilon_{kab} .
\end{equation} 
This is color-flavor locked (CFL) phase. The gauge symmetry  
and global chiral symmetry are broken according to the scheme
$SU(3)_c \times SU(3)_L\times SU(3)_R\rightarrow SU(3)_{L+R+c}$.
The baryon $U(1)$ symmetry is also broken. All quarks are gapped
in Cooper pairs with different colors and flavors. The lowest excitations
are nine Nambu-Goldstone bosons related to the spontaneous symmetry
breaking of global symmetries. There is also additional Nambu-Goldstone
boson connected to the breaking of restored axial symmetry $U_A(1)$.
The value of the gap is model dependent of the order of the 2SC gap but usualy 
slightly smaller.

The gap matrix can be written in general as:
\begin{equation}
\label{matrix}
\Delta^{ab}_{ij} =\left(
\begin{array}{ccccccccc}
0 & \Delta_{ud} & \Delta_{us} &  & & & & & \\
\Delta_{ud} & 0 & \Delta_{ds} &  & & & & & \\
\Delta_{us} & \Delta_{ds} & 0 &  & & & & & \\
 & & & 0 & -\Delta_{ud} &  & & & \\
 & & & -\Delta_{ud} & 0 &  & & & \\
 & & & & & 0 & -\Delta_{us} &  & \\
 & & & & & -\Delta_{us} & 0 &  & \\
 & & & & & & & 0 & -\Delta_{ds} \\
 & & & & & & & -\Delta_{ds} & 0 
\end{array}\right)
\end{equation}
in the basis 
\begin{equation}
(u_{red},d_{green},s_{blue},d_{red},u_{green},s_{red},u_{blue},s_{green},d_{blue}),
\end{equation}
where:
\begin{eqnarray}
\label{gaps}
\Delta_{us} = \Delta_{ds} = 0,\;\Delta_{ud} = \tilde{\Delta }\;\;\;\mbox{for 2SC}\\
\Delta_{us} = \Delta_{ds} = \Delta_{ud} = \Delta ,\;\;\;\;\;\mbox{for CFL}
\end{eqnarray}
From the gap matrix structre one can easily recognize the 
$(u_{red},d_{green}), (u_{green},u_{red})$ pairings in 2SC phase.
The quasiparticle excitations are two doublet representations of the remaining 
unbroken symmetry $SU(2)_L\times SU(2)_c$ of given chirality $L$. The second
similar set exists for oposite chirality. The unpaird quarks
are singlets under color rotations $SU(2)_c$ and doublets under
chiral symmetries. All of the excitations have the same gap $\tilde{\Delta }$.
The detail wave-functions of quasiparticle are given in Appendix B.
For the CFL phase we have pairings: $(d_{red},u_{green}), (s_{red},u_{blue}),
(s_{green},d_{blue})$ similar to the 2SC and the combination of
$(u_{red},d_{green},s_{blue})$ which is essentialy new for the CFL phase.
After diagonalization of the matrix we can find 8 excitations with
the gap $\Delta $ and one excitation with the gap $2\Delta $. These are
octet and singlet representations of the unbroken $SU(3)_{L+R+c}$ symmetry.
The detail calculations are given in Appendix C.

\section{Andreev reflection in superconducting QCD}

In this section, let us consider the Andreev reflection at the interface
consisting of two different superconductors in QCD, that is, the 2SC/CFL
interface since we find it is the most general case among the possible interfaces.
As is obvious from the expression of the gap matrix (21) with the conditions
(23) and (24), there are three kinds of possibilities of the quark scatterings at
this interface.

\begin{itemize}
\item $(s_{red},u_{blue})$ and/or $(s_{green},d_{blue})$: in this case, the Andreev reflection
is similar to QGP/2SC because strange quark as well as blue up and down
quarks are unpaired in the 2SC phase. This interface has already been studied \cite{sad}.
\item $(d_{red},u_{green})$: in this case, the reflection is 2SC/2SC' - like because
the gap of each phase is, in general, different. 
\item $(u_{red},d_{green},s_{blue})$:
in this case, the reflection is a mixture of (QGP-2SC)/CFL interface. This happens
because in the 2SC phase the $u_{red},d_{green}$ quarks are paired whereas 
strange quarks remain free.
\end{itemize}
Since the last two cases are specific to
the 2SC/CFL interface and have not been studied so far, let us restrict our consideration 
into those cases. The comparison to other possibilities would be given at the end
of this section.

Let us start with the 2SC/2SC' case. Physical setup we are interested in
is that the interface is placed at $z=0$ in space. For $z<0$, we have 2SC phase and for $z>0$,
CFL phase and we provide the boundary condition that matches the wavefunctions at $z=0$.

For $z<0$, the wavefunction takes the form as (below we put $m=0$ for simplicity)
\begin{eqnarray}
\label{2scwave1}
\Psi_< (z) = \left(
  \begin{array}{c}
    e^{i\frac{\delta_s}{2}}\sqrt{\frac{E+\lambda}{2E}}\varphi^{u}_{\uparrow R}   \\
    e^{-i\frac{\delta_s}{2}}\sqrt{\frac{E-\lambda}{2E}}h^{d\dagger{\rm T}}_{\downarrow L}   \\  \end{array}
\right)e^{ik_1z-iEt} \\\nonumber
+A\left(
  \begin{array}{c}
    e^{i\frac{\delta_s}{2}}\sqrt{\frac{E+\lambda}{2E}}\varphi^{u}_{\uparrow R}   \\
    e^{-i\frac{\delta_s}{2}}\sqrt{\frac{E-\lambda}{2E}}h^{d\dagger{\rm T}}_{\downarrow L}   \\  \end{array}
\right)e^{-ik_1 z-iEt}+
B\left(
  \begin{array}{c}
    e^{i\frac{\delta_s}{2}}\sqrt{\frac{E-\lambda}{2E}}\varphi^{u}_{\uparrow R}   \\
    e^{-i\frac{\delta_s}{2}}\sqrt{\frac{E+\lambda}{2E}}h^{d\dagger{\rm T}}_{\downarrow L}   \\  \end{array}
\right)e^{-ik_2 z-iEt},
\end{eqnarray}
while for $z>0$,
\begin{eqnarray}
\Psi_>(z) = C\left(
  \begin{array}{c}
    e^{i\frac{\delta_c}{2}}\sqrt{\frac{E+\xi}{2E}}\varphi^{u}_{\uparrow R}   \\
    e^{-i\frac{\delta_c}{2}}\sqrt{\frac{E-\xi}{2E}}h^{d\dagger{\rm T}}_{\downarrow L}   \\  \end{array}
\right)e^{ip_1 z-iEt}+
D\left(
  \begin{array}{c}
    e^{i\frac{\delta_c}{2}}\sqrt{\frac{E-\xi}{2E}}\varphi^{u}_{\uparrow R}  \\
    e^{-i\frac{\delta_c}{2}}\sqrt{\frac{E+\xi}{2E}}h^{d\dagger{\rm T}}_{\downarrow L}   \\  \end{array}
\right)e^{ip_2 z-iEt},
\label{2scwave2}
\end{eqnarray}
where $k_1 = \mu+\lambda, k_2=-\mu+\lambda, p_1=\mu+\xi$ and $p_2=-\mu+\xi$.
$\xi \equiv \sqrt{E^2-\Delta^2}$ and $\lambda \equiv \sqrt{E^2-\tilde{\Delta}^2}$.
$\tilde{\Delta}$ is the gap in the 2SC phase. The angles $\delta_s $ and $\delta_c$
are phases of the gap parameters in 2SC and CFL superconductors respectively.
The bispinors $\varphi_R ,h_L$ are desribed in Appendix A.

By matching these wavefunctions at $z=0$, we find the results as follows;
\begin{eqnarray}
A &=& 0, \nonumber \\
B &=& \frac{\sqrt{(E-\xi)(E+\lambda)}e^{i\alpha }-\sqrt{(E+\xi)(E-\lambda)}}
{\sqrt{(E+\xi)(E+\lambda)}-\sqrt{(E-\xi)(E-\lambda)e^{i\alpha}}}+O(1/\mu), \nonumber \\
C &=& \frac{2\lambda e^{i\frac{\alpha}{2}}}{\sqrt{(E+\xi)(E+\lambda)}-\sqrt{(E-\xi)(E-\lambda)e^{i\alpha}}}+O(1/\mu), \nonumber \\
D &=& 0, 
\label{result1}
\end{eqnarray}
where $\alpha = \delta_s - \delta_c$ is a phase difference of the gaps crossing the interface.
Note here that the coefficients $A$ and $D$ exactly vanish in the massless limit.
First of all, the transition and reflection coefficients, when $E > \Delta, \tilde{\Delta}$,  
are given by the formulae:
\begin{eqnarray}
T=\frac{\xi}{\lambda}|C|^2 &=& \frac{4\lambda\xi}{(E+\xi )(E+\lambda )}
\frac{1}{1+r^2-2r\cos\alpha }. \nonumber \\
R = 1-T
\label{b&c}
\end{eqnarray}
where 
\begin{equation}
r=\sqrt{\frac{(E-\lambda )(E-\xi )}{(E+\lambda )(E+\xi )}}
\end{equation}
whereas for $\Delta < E <\tilde{\Delta}$, $T=0$ and of course $R=1$.
The conserved current is defined as 
$j = \psi_>^{\dagger}\vec{\alpha}\psi_>=\psi_<^{\dagger}\vec{\alpha}\psi_<$
and coincides at both sides of the interface:
\begin{eqnarray}
j=
\left\{
  \begin{array}{cc}
 0 & {\rm for} \quad  \Delta >E>\tilde{\Delta}  \\ 
2\mu\frac{\xi}{E}\frac{4\lambda^2}{(E+\xi )(E+\lambda )(1+r^2-2r\cos\alpha )} 
& {\rm for} \quad E>\Delta,\tilde{\Delta}     
  \end{array}
\right.
\label{current}
\end{eqnarray}
The coefficients and currents depend on the phase difference $\alpha $ of the gap parameters.
We suggest that this combination is a gauge invariant quantity. Indeed in the
case of $U(1)$ superconductors $\alpha $ variable is responsible for many
interesting physical phenomena like, for example, the Josephson effect.

Let us move to the case (QGP - 2SC)/CFL interface (the mixed case).
In this case, the basis of the wavefunction must be 
\begin{eqnarray}
\left(\begin{array}{c} 
\label{basis}
\psi^{u}_{red}\\ \psi^{d}_{green}\\ \psi^{s}_{blue}\\ 
\psi^{u\dagger{\rm T}}_{red}\\ \psi^{d\dagger{\rm T}}_{green}\\ \psi^{s\dagger{\rm T}}_{blue}\\ 
\end{array}\right)
\end{eqnarray}
For $z>0$, we can use the equation (\ref{wf_cfl}) from Appendix C.
In the 2SC side the wavefunction takes the form:
\begin{eqnarray}
\Psi_<(z) &\equiv& 
\left(
  \begin{array}{c}
  e^{i\frac{\delta_s}{2}}\sqrt{\frac{E+\lambda}{2E}}\varphi^{u}_{\uparrow R}     \\
  0     \\
  0     \\
  0     \\
  e^{-i\frac{\delta_s}{2}}\sqrt{\frac{E-\lambda}{2E}}h^{d\dagger{\rm T}}_{\downarrow L}     \\
  0     \\
  \end{array}
\right)e^{ik_1z}
+ A\left(
  \begin{array}{c}
   e^{i\frac{\delta_s}{2}}\sqrt{\frac{E+\lambda}{2E}}\varphi^{u}_{\uparrow R}    \\
   0    \\
   0    \\
   0    \\
   e^{-i\frac{\delta_s}{2}}\sqrt{\frac{E-\lambda}{2E}}h^{d\dagger{\rm T}}_{\downarrow L}     \\
   0    \\
  \end{array}
\right)e^{-ik_1 z}+
B\left(
  \begin{array}{c}
   e^{i\frac{\delta_s}{2}}\sqrt{\frac{E-\lambda}{2E}}\varphi^{u}_{\uparrow R}     \\
  0     \\
  0     \\
  0     \\
  e^{-i\frac{\delta_s}{2}}\sqrt{\frac{E+\lambda}{2E}}h^{d\dagger{\rm T}}_{\downarrow L}      \\
  0     \\
  \end{array}
\right)e^{-ik_2 z} \nonumber \\
&+& C\left(
  \begin{array}{c}
   0    \\
  e^{i\frac{\delta_s}{2}}\sqrt{\frac{E+\lambda}{2E}}\varphi^{d}_{\uparrow R}     \\
   0    \\
   e^{-i\frac{\delta_s}{2}}\sqrt{\frac{E-\lambda}{2E}}h^{u\dagger{\rm T}}_{\downarrow L}     \\
   0    \\
   0    \\
  \end{array}
\right)e^{-ik_1 z} +
   D\left(
  \begin{array}{c}
   0    \\
   e^{i\frac{\delta_s}{2}}\sqrt{\frac{E-\lambda}{2E}}\varphi^{d}_{\uparrow R}    \\
  0     \\
  e^{-i\frac{\delta_s}{2}}\sqrt{\frac{E+\lambda}{2E}}h^{u\dagger{\rm T}}_{\downarrow L}     \\
   0    \\
  0     \\
  \end{array}
\right)e^{-ik_2 z}+
F\left(
  \begin{array}{c}
   0    \\
   0    \\
   \varphi^{s}_{\uparrow R}    \\
   0    \\
   0    \\
   0    \\
  \end{array}
\right)e^{-ik_3 z}+
G\left(
  \begin{array}{c}
   0    \\

   0    \\
  0     \\
  0     \\
  0     \\
  h^{s\dagger{\rm T}}_{\downarrow L}     \\
  \end{array}
\right)e^{-ik_4 z},
\label{2scwave3}
\end{eqnarray}
The first wavefunction describes the incoming quasiparticle, second and third the reflected
quasiparticles of the same kind. The next two wavefunctions are quasiparticles
of the second kind whereas the last two describe the posssibility of the reflection of
the unpaired strange quark. The momenta of the strange quark and hole are 
$k_3=\mu+E$ and $k_4=-\mu+E$, the other
were already given previously. On the other hand, for $z>0$, we have 
the wavefunction given by the formula (\ref{wf_cfl_qgp}) in Appendix C.
In the massless limit we consider here
the momenta in the wave function (\ref{wf_cfl_qgp}) are given by 
$q_1 = \mu+\xi $, $q_2=-\mu+\xi $, $p_1=\mu+\zeta$ and $p_2=-\mu+\zeta$.
Solving the boundary condition which connects $\Psi_<(z)$ with $\Psi_>(z)$, we find the following
results (the phase $\exp(i\delta_s/2)$ was absorbed into the redefinition of
constants $G, H, K$ and $N$):
\begin{eqnarray}
B = -l + \frac{2e^{i\alpha} (e^{i\alpha }l x (x-2z)+x+z)\lambda }{(1-e^{i\alpha} lx)(3+e^{i\alpha } x (x-2 z))(E+\lambda )} \\\nonumber
D = - \frac{2e^{i\alpha }(2x-z)\lambda }{(1-e^{i\alpha} lx)(3+e^{i\alpha } l (x-2 z))(E+\lambda )} \\\nonumber
G = \frac{\lambda }{E}\sqrt{\frac{2 E}{E+\lambda}}\frac{(x+z)e^{-i\delta_{c}}}{3+e^{i\alpha } l (x-2 z)} \\\nonumber
H = - \frac{\lambda }{E}\sqrt{\frac{2 E}{E+\lambda}}\frac{x(1+l(x-z)e^{i\alpha})e^{-i\delta_{c}}}{(1-e^{i\alpha} lx)(3+e^{i\alpha } l (x-2 z))} \\\nonumber
K = \frac{\lambda }{E}\sqrt{\frac{2 E}{E+\lambda}}\frac{xe^{-i\delta_{c}}}{3+e^{i\alpha } l (x-2 z)} \\\nonumber
N = \frac{\lambda }{E}\sqrt{\frac{2 E}{E+\lambda}}\frac{ze^{-i\delta_{c}}}{3+e^{i\alpha } l (x-2 z)} \\\nonumber
\end{eqnarray}
where 
\begin{equation}
x = \sqrt{\frac{E-\xi}{E+\xi}},\; l = \sqrt{\frac{E-\lambda}{E+\lambda}},\;x = \sqrt{\frac{E-\zeta}{E+\zeta}},
\end{equation}
Other coefficients ($A, C, D, F, J, L$ and $P$) vanish.
Let us notice that the modulus square of the above coefficients depends only
on the phase difference $\alpha $ crossing the interface. Similarly to the
case of 2SC/2SC' junction this quantity we consider is gauge independent.

The probability current for $z<0$ and $E > \tilde{\Delta}, 2\Delta $ is
\begin{equation}
j \equiv \Psi_<^{\dagger}\vec{\alpha}\Psi_<
= 2\mu[\frac{\xi}{E}(1-|B|^2-|D|^2)-|G|^2].
\label{current2}
\end{equation}
The analytical result is not given by short expression thus we rather show
the current dependence on the energy in the Fig. 1 for generic set of parameters.
Let us notice that the current starts at energy $E=\Delta $ and then rises
linearly up to the point $E=2\Delta $ where there is a jump. This behaviour is expected
because below the gap there is no probability current and at the energy twice the gap
additional current appears from the singlet excitation in the CFL phase. 

\begin{figure}
\centerline{\epsfxsize=7 cm \epsfbox{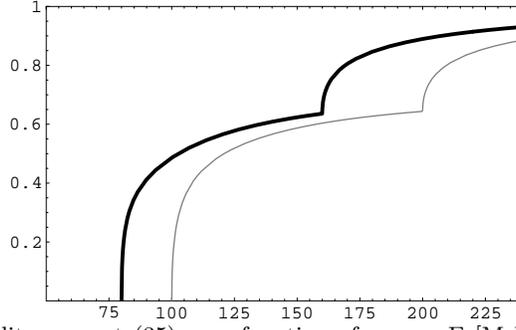}}
\caption{Dependence of the probability current (\ref{current2}) as a function of energy E [MeV] for
two sets of parameters: $\Delta = 80$ MeV, $\tilde\Delta = 60$ MeV (black curve) and
$\Delta = 100$ MeV, $\tilde\Delta = 60$ MeV (gray curve).}
\end{figure}

\bigskip

For comparison let us show the result for transition and reflection coefficients
in the QGP/CFL scattering:
\begin{eqnarray}
T^3 = \left\{\begin{array}{cc}
0 & \;\;\;\mbox{for}\, E<|\Delta | \\
\frac{\xi }{E+\xi } & \;\;\;\mbox{for}\, E>|\Delta | 
\end{array}\right.\\\nonumber
T^8 = \left\{\begin{array}{cc}
0 & \;\;\;\mbox{for}\, E<|\Delta | \\
\frac{\xi }{3(E+\xi )} & \;\;\;\mbox{for}\, E>|\Delta | 
\end{array}\right.\\\nonumber
T^9 = \left\{\begin{array}{cc}
0 & \;\;\;\mbox{for}\, E<2|\Delta | \\
\frac{2\zeta }{3(E+\zeta )} & \;\;\;\mbox{for}\, E>2|\Delta | 
\end{array}\right.\\\nonumber
\end{eqnarray}
where $T^A$ are transition coefficients for quasiparicles in the CFL basis
described by the wavefunctions (\ref{wf_cfl}) given in the Appendix C. 
Let us note that because of the energy conservation there is always only 
one hole reflected in the QGP phase:
\begin{eqnarray}
\label{refl}
R^{u}_{red} = \left\{
\begin{array}{ll}
\frac{5}{9}-\frac{2}{9|\Delta |^2}[(E^2+|\zeta ||\xi |)\cos (2\delta )-
E(|\zeta |-|\xi |)\sin (2\delta )] & \;\mbox{for} E <|\Delta | \\
\frac{1}{9}+\frac{4 (E-\xi )}{9 (E+\xi )}-\frac{2}{9(E+\xi )}
[E\cos (2\delta )-|\zeta |\sin (\delta )] & \;\mbox{for} |\Delta |<E<2|\Delta | \\
\frac{1}{9}\frac{E-\zeta }{E+\zeta }+\frac{4}{9}\frac{E-\xi }{E+\xi }-
\frac{2}{9}\frac{E-\zeta }{E+\xi }\cos (2\delta ) & \;\mbox{for} E>2|\Delta | 
\end{array}\right. \\\nonumber
R^{d}_{green} = R^{s}_{blue} = \left\{
\begin{array}{ll}
\frac{2}{9}+\frac{1}{9|\Delta |^2}[(E^2+|\zeta ||\xi |)\cos (2\delta )-
E(|\zeta |-|\xi |)\sin (2\delta )] & \;\mbox{for} E <2|\Delta | \\
\frac{1}{9}+\frac{E-\xi }{9 (E+\xi )}+\frac{1}{9(E+\xi )}
[E\cos (2\delta )-|\zeta |\sin (\delta )] & \;\mbox{for} |\Delta |<E<2|\Delta | \\
\frac{1}{9}\frac{E-\zeta }{E+\zeta }+\frac{1}{9}\frac{E-\xi }{E+\xi }+
\frac{1}{9}\frac{E-\zeta }{E+\xi }\cos (2\delta ) & \;\mbox{for} E>2|\Delta | 
\end{array}\right. \\\nonumber
\end{eqnarray}
where $|\zeta |=\sqrt{4|\Delta |^2-E^2}, |\xi |=\sqrt{|\Delta |^2-E^2}$.
Reflection and transition coefficients sum up to unity for any value
of energy $E$ of incoming particle. The coefficients (\ref{refl}) depends
on the phase $\delta $ of the gap parameter. However this single phase is
a gauge dependent quantity. Thus to get the physical answer we have to
avarage over the phase angle. The sinus and cosinus avarage to zero giving
the final answer in the form:
\begin{eqnarray}
\label{refl2}
R^{u}_{red} = \left\{
\begin{array}{ll}
\frac{5}{9} & \;\mbox{for} E <|\Delta | \\
\frac{1}{9}+\frac{4 (E-\xi )}{9 (E+\xi )} & \;\mbox{for} |\Delta |<E<2|\Delta | \\
\frac{1}{9}\frac{E-\zeta }{E+\zeta }+\frac{4}{9}\frac{E-\xi }{E+\xi } & \;\mbox{for} E>2|\Delta | 
\end{array}\right. \\\nonumber
R^{d}_{green} = R^{s}_{blue} = \left\{
\begin{array}{ll}
\frac{2}{9} & \;\mbox{for} E <2|\Delta | \\
\frac{1}{9}+\frac{E-\xi }{9 (E+\xi )} & \;\mbox{for} |\Delta |<E<2|\Delta | \\
\frac{1}{9}\frac{E-\zeta }{E+\zeta }+\frac{1}{9}\frac{E-\xi }{E+\xi }
& \;\mbox{for} E>2|\Delta | 
\end{array}\right. \\\nonumber
\end{eqnarray}

For the case of QGP/2SC scattering the results are the same as given
by equations (\ref{nr_current},\ref{nr_rt}). The only difference is that instead
of Fermi velocity $v_F$ in the probability current (\ref{nr_current}) we have relativistic
expression $2p_F$ (the Fermi momentum $p_F$ is essentialy equal to
quark chemical potential $\mu $ in our approximation). Let us notice that
the reflection and transition coefficients are exactly the same like
in the non-relativistic case.

\section{Conclusions}

In this paper we consider the general structure of the Andreev reflection
between two superconductors 2SC and CFL in the high density QCD. We also give
the review of the Andreev reflection between QGP/2SC and
QGP/CFL phases as well as in condensed matter systems.

The essence of the AR stems from the peculiar behaviour of particles which
hit the interface. If the energy of the incoming particle from
the conductor side is below the energy gap in superconductor then the hole is reflected from the
interface. The energy and momentum\footnote{The momentum conservation is violated
at the $O(1/E_F$) level and can be neglected in the first approximation.} are conserved,
however, there is an apparent violation of the charge conservation. This last effect
stems from the fact that the superconductor serves as an infinite suply
of Cooper pairs. At the microscopic level charge is conserved, of course.
The Andreev reflection can be understood as follows: the incoming particle
takes another particle from the Fermi sea and creates a Cooper
pair which dissolves in the condensate. Then the hole is left
in the conductor with the appropriate kinematics constaraind
by energy and momentum conservation.

This pattern is exactly repeated in QCD superconductors.
The differences come from the more complicated structure of
Cooper pairing of quarks. There are more quantum numbers describing QGP phase
as well as the quasiparticle structure
in QCD superconductors are richer. The treatment of the
Andreev reflection processes are similar to the case of nonrelativistic 
physics, however, more attention has to be taken for the gauge
invariance of the final results. It appears that the
probability currents are gauge independent quantities
but the transition $T$ and the reflection $R$ coefficients 
require more attention. Our effective hamiltonian (\ref{hamiltonian_qcd})
is not gauge invariant, nevertheless one can trace the gauge behaviour by
tracing the dependence of the final result on the phase $\delta $ of the gap parameter.
Obviously the phase $\delta $ itself is not gauge invariant quantity.
In the QGP/2SC case the coefficents $T$ and $R$ are gauge invariant,
however, this is not true for the QGP/CFL interface (\ref{refl}). Let us first notice
that one should expect the gauge invariant answers for $T$ and $R$ coefficents
because the reflected holes can be distinguished by their flavor quantum number
(not a color). We suggest that the avaraging of the final result over the phase
$\delta$ rebuilts the gauge invariant answer. This is reasonable. Indeed 
to perform the Andreev reflection experiment one has to prepare the ensemble of the
QGP/CFL interfaces. Then each sample has its own unphysical gap phase.
Then performing the experiment on the unsamble is equivalent to
avarage over the phase $\delta $. Finally in the case of 2SC/CFL interface
the physical quantities depend on the phase difference $\alpha $ crossing the boundary.
This is actualy expected result because, similarly to $U(1)$ superconductors, the
angle $\alpha $ is a gauge independent variable.

The Andreev reflection has many interesting physical effects
in condensed matter systems \cite{andreev,benistant,blonder}.
In the high density QCD situation is more complicated. The only
known place one can expect the color superconductors are the cores
of protoneutron or Neutron Stars. Any subtle effets of Andreev reflection that happen inside
these objects far away from our laboratories may be never observed.
However there is at least one possibility one can imagine. The
Andreev reflection affects the transport properties at the interfaces. If the
protoneutron stars go through the first order phase transitions between
QGP or/and 2SC or/and CFL phases during their history cooling then one can expect
the mixed phase to be present in the core of the stars. In this situation the Andreev reflection
influences the dynamics of the bubbles grow between different phases. These
in turn influence the neutrino emmission. Thus the time dependence of
neutrino luminosity from supernovae can carry the information of the
Andreev processes which took place inside the protoneutron stars.
These intriguing possibilities require more attention.

\bigskip

{\bf Acknowledgement}. M.S was supported by a fellowship from the Foundation for
Polish Science. M.S. was also supported in part by
Polish State Committee for Scientific Reasearch, grant no. 2P 03B 094 19.

\bigskip

\begin{center}
{\bf APPENDIX A - Particles, holes and Dirac algebra}
\end{center}

Let us consider the equations of motion for free quarks at chemical
potential $\mu $. 
\begin{eqnarray}
\label{eom_free}
i\dot{\psi}(t,\vec{r}) &=& (-i\vec{\alpha}\cdot\vec{\nabla}+m\gamma_0-\mu)\psi (t,\vec{r}), \nonumber \\
i\dot{\psi}^{\dagger}(t,\vec{r})
&=& -i\vec{\nabla}\psi^{\dagger}(t,\vec{r})\cdot
\vec{\alpha}-\psi^{\dagger}(t,\vec{r})(m\gamma_0-\mu).
\end{eqnarray}
Remembering that relativistic effects are suppressed at high density the Dirac 
field operator $\psi (t,\vec{r})$ can have a one-particle interpretation
of annihilating a fermion of given spin at the point $(t,\vec{r})$ in space-time. We can
also change the operator equation (\ref{eom_free}) into the wavefunction
equation. The first equation of (\ref{eom_free}) describes the particles and the
second one the holes. Using the decomposition:
\begin{eqnarray}
\label{decomp}
\psi (t, \vec{r}) = 
\sum_{r}\alpha_r\varphi_{r,R}(\vec{q})\exp(i\vec{q}\cdot\vec{r}-i\epsilon t),\\\nonumber
\psi^{\dagger}(t, \vec{r}) 
= \sum_{r}\gamma_r^* h^{\dagger}_{r,L}(-\vec{q})\exp(i\vec{q}\cdot\vec{r}-i\epsilon t),
\end{eqnarray}
one can easily solve the equations (\ref{eom_free}). To finish this
let us define the bispinors $\varphi_{r,R}(\vec{k})$ and
$\varphi_{r,L}(\vec{k})$ through the equations:
\begin{eqnarray}
(\alpha\cdot\vec{k}+m\gamma_0-\mu )\varphi_{r,R}(\vec{k}) = 
\epsilon\varphi_{r,R}(\vec{k})\\\nonumber
\varphi^{\dagger }_{r,L}(\vec{k})(\alpha\cdot\vec{k}+m\gamma_0-\mu )= 
\epsilon\varphi^{\dagger}_{r,L}(\vec{k})
\end{eqnarray}
where $r$ describes the spin and momentum $\vec{k}$ can be in
general complex vector. From this reason one has to distinguish
between the right- and left-handed eigenvectors, which are
denoted by the capital letters $L,R$. For complex $\vec{k}$ 
$\varphi^{\dagger }_{r,L}$ is not hermitian conjugate to $\varphi_{r,R}$.
The solutions of the equations take the form:
\begin{eqnarray}
\varphi_{r,R}(\vec{k})=\left( 
\begin{array}{c}
\sqrt{m+\epsilon +\mu} \chi_r \\
\frac{\vec{\sigma }\cdot\vec{k}}{\sqrt{m+\epsilon +\mu }}\chi_r \\
\end{array}
\right) \\\nonumber
\varphi^\dagger_{r,L}(\vec{k})=\left(\bar{\chi}_{r}^\dagger\sqrt{m + \epsilon + \mu},
\bar{\chi}_{r}^\dagger\frac{\vec{\sigma }\cdot\vec{k}}{\sqrt{m+\epsilon +\mu }}\right)
\end{eqnarray}
where $\chi_r ,\bar{\chi}_{r}^\dagger $ are spinors and where 
$\epsilon = \sqrt{\vec{k}^2+m^2}-\mu $\footnote{There is another
solution with $\epsilon = -\sqrt{\vec{k}^2+m^2}-\mu $ which is not interesting
for our purposes}.
The spinors can be defined in the helicity basis:
\begin{eqnarray}
\vec{\sigma }\cdot\vec{k}\chi_{\uparrow ,\downarrow }= 
\pm k\chi_{\uparrow ,\downarrow }\\\nonumber
\bar{\chi}_{\uparrow ,\downarrow }^\dagger\vec{\sigma }\cdot\vec{k}
= \pm k\bar{\chi}_{\uparrow ,\downarrow }^\dagger
\end{eqnarray} 
where $k=\sqrt{\vec{k}^2}$.
Let us also define the additional bispinors:
\begin{eqnarray}
h^\dagger_{r,L}(\vec{k})(\alpha\cdot\vec{k}- m\gamma_0 + \mu ) = 
\bar{\epsilon } h^\dagger_{r,L}(-\vec{k})\\\nonumber
(\alpha\cdot\vec{k}- m\gamma_0 + \mu )h_{r,R}(-\vec{k}) = \bar{\epsilon }h_{r,R}(-\vec{k})
\end{eqnarray}
where $\bar{\epsilon }=-\epsilon $ and bispinors are given by formulae: 
\begin{eqnarray}
h_{r,R}(-\vec{k})=\left( 
\begin{array}{c}
\sqrt{m-\epsilon +\mu} \chi_r \\
\frac{\vec{\sigma }\cdot\vec{k}}{\sqrt{m-\epsilon +\mu }}\chi_r \\
\end{array}
\right) \\\nonumber
h^\dagger_{r,L}(-\vec{k})=\left({\bar{\chi}_r}^\dagger\sqrt{m - \epsilon + \mu},
-{\bar{\chi}_r}^\dagger\frac{\vec{\sigma }\cdot\vec{k}}{\sqrt{m-\epsilon +\mu }}\right)
\end{eqnarray}
The bispinors defined above fulfill simple algebraic relations
which are useful in the calculations:
\begin{eqnarray}
\varphi_{r,L}^\dagger\varphi_{s,R}=h_{r,L}^\dagger h_{s,R}=2\sqrt{k^2+m^2}\delta_{rs}\\\nonumber
\varphi_{s,L}^\dagger C\gamma_5 h_{r,L}^{\dagger\,T} h_{s,R} =
\varphi_{s,R}^{T} C\gamma_5 h_{r,R} = 2\sqrt{k^2+m^2}
\left\{
\begin{array}{lr}
-1 & s=\uparrow\,r=\downarrow \\
1 & s=\downarrow\, r=\uparrow 
\end{array}
\right.
\end{eqnarray}

The bispinors defined above together with (\ref{decomp}) are
solutions of (\ref{eom_free}) and have simple physical meaning.
The wavefunction:
\begin{equation}
\psi (t,\vec{r}) = \varphi_{\uparrow\, R}(\vec{k})\exp{(-i\epsilon t+i\vec{k}\cdot\vec{r})}
\end{equation}
describes the particle of spin projection up, velocity $\vec{v} = \frac{\vec{k}}{E}$, 
where $E = \sqrt{{\vec{k}}^2+m^2}$ and energy $\epsilon $ above
the Fermi Sea. From the other hand the wavefunction:
\begin{equation}
\psi^\dagger (t,\vec{r}) = 
h^{\dagger}_{\downarrow\, L}(-\vec{k})\exp{(-i\epsilon t+i\vec{k}\cdot\vec{r})}
\end{equation}
describes the hole of spin projection down, velocity $-\vec{v}$, 
and energy $\epsilon$ below the Fermi Sea.

\begin{center}
{\bf APPENDIX B - 2SC quasiparticles}
\end{center}

From the hamiltonian (\ref{hamiltonian_qcd}) and the gap structure
(\ref{matrix}) for 2SC phase (\ref{gaps}), using usual commutation relations
for fermi fields, one can derive the equation of motion:
\begin{eqnarray}
\label{eom_2SC}
i\dot{\psi}^{u}_{red} = (-i\vec{\alpha}\cdot\vec{\nabla}+m\gamma_0-\mu)\psi^{u}_{red}
-\Delta C\gamma_5 \psi^{d*}_{green}, \nonumber \\
i\dot{\psi}^{d}_{green} = (-i\vec{\alpha}\cdot\vec{\nabla}+m\gamma_0-\mu)\psi^{d}_{green}
-\Delta C\gamma_5 \psi^{u*}_{red}
\end{eqnarray}
where we exchange the field operators by the wave functions similarly
like in the nonrelativistic case. This can be done because the presence of the Fermi sea suppresses
the contribution of antiquarks. There is similar set of equations for another $(u_{green},d_{red})$
pair. It is convenient to use the plane wave decomposition:
\begin{eqnarray}
\label{decomp_2SC}
\psi^u_{red}(t, \vec{r}) = \sum_{r}\alpha_r\varphi^{u}_{r,R}(\vec{q})\exp(i\vec{q}\cdot\vec{r}-iEt), \nonumber \\
\psi^d_{green}(t, \vec{r}) = \sum_{r}\beta_r \varphi^{d}_{r,R}(\vec{q})\exp(i\vec{q}\cdot\vec{r}-iEt)
\end{eqnarray}
where the bispinors $\varphi$ and $h$ are defined in Appendix A. $\alpha_r, \beta_r$ 
are some constants. The subscripts $u, d$ and $s$ 
describe flavor and color of quarks. Plugging (\ref{decomp_2SC}) into (\ref{eom_2SC}),
using the relations given in Appendix A one can find the wave functions \cite{sad}:
\begin{eqnarray}
\label{wf_2SC}
\Psi(t, \vec{r}) \equiv \left(\begin{array}{c} 
\psi^{u}_{red}\\ 
\psi^{d\dagger{\rm T}}_{green}
\end{array}\right) =
\left(
  \begin{array}{c}
    \exp (i\delta /2)\sqrt{\frac{E+\lambda}{2E}}(C\varphi^{u}_{\uparrow R}-C^\prime\varphi^{u}_{\downarrow R})   \\
    \exp (-i\delta /2)\sqrt{\frac{E-\lambda}{2E}}(C h^{d\dagger{\rm T}}_{\downarrow L}+C^\prime h^{d\dagger{\rm T}}_{\uparrow L})   \\  
\end{array}
\right)e^{ip_1 z-iEt}+ \\\nonumber
\left(
  \begin{array}{c}
    \exp (i\delta /2)\sqrt{\frac{E-\lambda}{2E}}(D\varphi^{u}_{\uparrow R}-D^\prime\varphi^{u}_{\downarrow R})  \\
    \exp (-i\delta /2)\sqrt{\frac{E+\lambda}{2E}}(D h^{d\dagger{\rm T}}_{\downarrow L}+D^\prime h^{d\dagger{\rm T}}_{\uparrow L})   \\  
\end{array}
\right)e^{ip_2 z-iEt},
\end{eqnarray}
where $\delta $ is a phase of the gap parameter, $p_{1,2} = \sqrt{(\mu\pm\lambda )^2-m^2}$,
$\lambda = \sqrt{E^2-|\tilde{\Delta}|^2}$ and $C, C^\prime , D, D^\prime $ are some constants. The first wavefunction
describes the particle-like excitation whereas the second one describes the hole-like
quasiparticle. This is quite similar to the description of the nonrelativistic
quasiparticles. Indeed if one compares (\ref{wf_2SC}) with the wavefunctions
(\ref{quasi}) the similarity is striking. It also occurs that the Andreev reflection
in the QGP/2SC case is a direct analog of the Andreev reflection at
the nonrelativistic conductor/superconductor junction.

The extension of the wave function (\ref{wf_2SC}) to the case
of the 2SC superconductor with 3 flavors is strightforward. The exact
expressions are given in Section V.

\begin{center}
{\bf APPENDIX C - CFL quasiparticles}
\end{center}

Let us now discuss the CFL quasiparticle waveufunctions for the
states $u_{red}, d_{green}, s_{blue}$.
The equations of motions that follow from (\ref{hamiltonian_qcd}) take the form: 
\begin{eqnarray}
i\dot{\psi}^{u}_{red} &=& (-i\vec{\alpha}\cdot\vec{\nabla}+m\gamma_0-\mu)\psi^{u}_{red}
-\Delta C\gamma_5 (\psi^{d*}_{green}+\psi^{s*}_{blue}), \nonumber \\
i\dot{\psi}^{d}_{green} &=& (-i\vec{\alpha}\cdot\vec{\nabla}+m\gamma_0-\mu)\psi^{d}_{green}
-\Delta C\gamma_5 (\psi^{u*}_{red}+\psi^{s*}_{blue}), \nonumber \\
i\dot{\psi}^{s}_{blue} &=& (-i\vec{\alpha}\cdot\vec{\nabla}+m\gamma_0-\mu)\psi^{s}_{blue}
-\Delta C\gamma_5 (\psi^{u*}_{red}+\psi^{d*}_{green}), \nonumber \\
i\dot{\psi}^{u\dagger}_{red} 
&=& -i\vec{\nabla}\psi^{u\dagger}_{red}\cdot\vec{\alpha}-\psi^{u\dagger}_{red}(m\gamma_0-\mu)
-\Delta^* (\psi^{d\rm{T}}_{green}+\psi^{s\rm{T}}_{blue})C\gamma_5, \nonumber \\
i\dot{\psi}^{d\dagger}_{green} 
&=& -i\vec{\nabla}\psi^{d\dagger}_{green}\cdot\vec{\alpha}-\psi^{d\dagger}_{green}(m\gamma_0-\mu)
-\Delta^* (\psi^{u\rm{T}}_{red}+\psi^{s\rm{T}}_{blue})C\gamma_5, \nonumber \\
i\dot{\psi}^{s\dagger}_{blue} 
&=& -i\vec{\nabla}\psi^{s\dagger}_{blue}\cdot\vec{\alpha}-\psi^{s\dagger}_{blue}(m\gamma_0-\mu)
-\Delta^* (\psi^{u\rm{T}}_{red}+\psi^{d\rm{T}}_{green})C\gamma_5.
\label{eom_cfl}
\end{eqnarray}

To find the quasiparticle wavefunctions for $\Delta = const$, 
it is convenient to use the following decompositions:
\begin{eqnarray}
\psi^u_{red}(t, \vec{r}) = \sum_{r}\alpha_r\varphi^{u}_{r,R}(\vec{q})\exp(i\vec{q}\cdot\vec{r}-iEt),
& & \quad
\psi^{u\dagger}_{red}(t, \vec{r}) 
= \sum_{r}\alpha_r^* h^{u\dagger}_{r,L}(-\vec{q})\exp(i\vec{q}\cdot\vec{r}-iEt),
\nonumber \\
\psi^d_{green}(t, \vec{r}) = \sum_{r}\beta_r \varphi^{d}_{r,R}(\vec{q})\exp(i\vec{q}\cdot\vec{r}-iEt),
& & \quad 
\psi^{d\dagger}_{green}(t, \vec{r}) 
= \sum_{r}\beta_r^* h^{u\dagger}_{r,L}(-\vec{q})\exp(i\vec{q}\cdot\vec{r}-iEt),
\nonumber \\
\psi^s_{blue}(t, \vec{r}) = \sum_{r}\gamma_r \varphi^{s}_{r,R}(\vec{q})\exp(i\vec{q}\cdot\vec{r}-iEt),
& & \quad
\psi^{s\dagger}_{blue}(t, \vec{r}) 
= \sum_{r}\gamma_r^* h^{u\dagger}_{r,L}(-\vec{q})\exp(i\vec{q}\cdot\vec{r}-iEt),
\label{expand}
\end{eqnarray}
where $\varphi$ and $h$ are defined in Appendix A. $\alpha_r, \beta_r$ 
and $\gamma_r$ are some constants. The subscripts $u, d$ and $s$ 
describe flavor and color of quarks in obvious way. 

Plugging (\ref{expand}) into (\ref{eom_cfl}), using the bispinor algebraic relations given
in Appendix A and assuming constant value of the gap parameter $\Delta$, one obtains the
wavefunction describing the quasiparticle excitations of given energy $E$ in the CFL phase:
\begin{eqnarray}
\label{wf_cfl_qgp}
\Psi(t, \vec{r}) \equiv \left(\begin{array}{c} 
\psi^{u}_{red}\\ \psi^{d}_{green}\\ \psi^{s}_{blue}\\ 
\psi^{u\dagger{\rm T}}_{red}\\ \psi^{d\dagger{\rm T}}_{green}\\ \psi^{s\dagger{\rm T}}_{blue}\\ 
\end{array}\right)
=\left[ H\left(\begin{array}{c} 
\frac{E+\xi}{\Delta^*}\varphi^{u}_{\uparrow R}\\ 
-\frac{E+\xi}{\Delta^*}\varphi^{d}_{\uparrow R}\\ 0\\ 
-h^{u\dagger{\rm T}}_{\downarrow L}\\ h^{d\dagger{\rm T}}_{\downarrow L}\\ 0\\ 
\end{array}\right)e^{i\vec{q}_1\cdot\vec{r}}
+ J\left(\begin{array}{c} 
\frac{E-\xi}{\Delta^*}\varphi^{u}_{\uparrow R}\\ 
-\frac{E-\xi}{\Delta^*}\varphi^{d}_{\uparrow R}\\ 0\\ 
-h^{u\dagger{\rm T}}_{\downarrow L}\\ h^{d\dagger{\rm T}}_{\downarrow L}\\ 0\\
 \end{array}\right)e^{i\vec{q}_2\cdot\vec{r}} 
+ K\left(\begin{array}{c} 
\frac{E+\xi}{\Delta^*}\varphi^{u}_{\uparrow R}\\ 0\\
-\frac{E+\xi}{\Delta^*}\varphi^{s}_{\uparrow R}\\  
-h^{u\dagger{\rm T}}_{\downarrow L}\\ 0\\ h^{s\dagger{\rm T}}_{\downarrow L}\\  
\end{array}\right)e^{i\vec{q}_1\cdot\vec{r}} +\right. \nonumber \\
\left. + L\left(\begin{array}{c} 
\frac{E-\xi}{\Delta^*}\varphi^{u}_{\uparrow R}\\ 0\\
-\frac{E-\xi}{\Delta^*}\varphi^{s}_{\uparrow R}\\  
-h^{u\dagger{\rm T}}_{\downarrow L}\\ 0\\ h^{s\dagger{\rm T}}_{\downarrow L}\\  
\end{array}\right)e^{i\vec{q}_2\cdot\vec{r}}
+ N\left(\begin{array}{c} 
\frac{2\Delta}{E-\zeta}\varphi^{u}_{\uparrow R}\\ 
\frac{2\Delta}{E-\zeta}\varphi^{d}_{\uparrow R}\\
\frac{2\Delta}{E-\zeta}\varphi^{s}_{\uparrow R}\\   
h^{u\dagger{\rm T}}_{\downarrow L}\\ 
h^{d\dagger{\rm T}}_{\downarrow L}\\ 
h^{s\dagger{\rm T}}_{\downarrow L}\\  \end{array}\right)e^{i\vec{p}_1\cdot\vec{r}}
+ P\left(\begin{array}{c} 
\frac{2\Delta}{E+\zeta}\varphi^{u}_{\uparrow R}\\ 
\frac{2\Delta}{E+\zeta}\varphi^{d}_{\uparrow R}\\
\frac{2\Delta}{E+\zeta}\varphi^{s}_{\uparrow R}\\   
h^{u\dagger{\rm T}}_{\downarrow L}\\ 
h^{d\dagger{\rm T}}_{\downarrow L}\\ 
h^{s\dagger{\rm T}}_{\downarrow L}\\  \end{array}\right)e^{i\vec{p}_2\cdot\vec{r}} 
\right] \exp{(-iEt)}
\end{eqnarray}
where $\xi = \sqrt{E^2-|\Delta |^2}$ and $\zeta = \sqrt{E^2-4|\Delta |^2}$.
$q_{1,2} = \sqrt{(\mu \pm \xi)^2-m^2}$ and $p_{1,2} = \sqrt{(\mu \pm \zeta)^2-m^2}$.
$H, J, K, L, N$ and $P$ are arbitrary constants. Similar expression is obtained
for the opposite spin content. 
Instead of using $u_{red}, d_{green}, s_{blue}$ basis one can use the CFL-basis:
\begin{equation}
\psi^{i}_{\alpha } = \sum_{A=1}^9 \frac{(\lambda^A)_{i\alpha }}{\sqrt{2}}\psi^A
\end{equation}
 in which the gap matrix (\ref{matrix}) is diagonal, where
\begin{equation}
\lambda^A,\,\,A=1,...,8\;\;\mbox{are Gell-Mann matrices and}\;\;
\lambda^9=\sqrt{\frac{2}{3}}\hat{1}
\end{equation}
 Explicitly:
\begin{eqnarray}
\label{cfl_expl}
\psi^3 = \frac{1}{\sqrt{2}}(\psi^{u}_{red} - \psi^{d}_{green})\\\nonumber
\psi^8 = \frac{1}{\sqrt{6}}(\psi^{u}_{red} + \psi^{d}_{green} - 2 \psi^{s}_{blue})\\\nonumber
\psi^9 = \frac{1}{\sqrt{6}}(\psi^{u}_{red} + \psi^{d}_{green} + \psi^{s}_{blue})
\end{eqnarray}
In the CFL-basis one can write quasiparticle wavefinctions as:
\begin{eqnarray}
\label{wf_cfl}
\psi^{9}_{CFL}\equiv
\left(\begin{array}{c}
\psi^{9}_{particle}\\
\psi^{9}_{hole}
\end{array}\right) = \sqrt{3}N
\left(\begin{array}{c}
\frac{2\Delta }{E-\zeta }\varphi_{\uparrow\, R} \\
h^{\dagger{\rm T}}_{\downarrow\, L}
\end{array}\right)e^{i\vec{p}_1\cdot\vec{r}} +
\sqrt{3}P
\left(\begin{array}{c}
\frac{2\Delta }{E+\zeta }\varphi_{\uparrow\, R} \\
h^{\dagger{\rm T}}_{\downarrow\, L}
\end{array}\right)e^{i\vec{p}_2\cdot\vec{r}} \\\nonumber
\psi^{8}_{CFL}\equiv
\left(\begin{array}{c}
\psi^{8}_{particle}\\
\psi^{8}_{hole}
\end{array}\right) = \sqrt{\frac{3}{2}}K
\left(\begin{array}{c}
\frac{E+\xi }{\Delta^\ast }\varphi_{\uparrow\, R} \\
-h^{\dagger{\rm T}}_{\downarrow\, L}
\end{array}\right)e^{i\vec{q}_1\cdot\vec{r}} +
\sqrt{\frac{3}{2}}L
\left(\begin{array}{c}
\frac{E-\xi }{\Delta^\ast }\varphi_{\uparrow\, R} \\
-h^{\dagger{\rm T}}_{\downarrow\, L}
\end{array}\right)e^{i\vec{q}_2\cdot\vec{r}} \\\nonumber
\psi^{3}_{CFL}\equiv
\left(\begin{array}{c}
\psi^{3}_{particle}\\
\psi^{3}_{hole}
\end{array}\right) = \frac{2H + K}{\sqrt{2}}
\left(\begin{array}{c}
\frac{E+\xi }{\Delta^\ast }\varphi_{\uparrow\, R} \\
-h^{\dagger{\rm T}}_{\downarrow\, L}
\end{array}\right)e^{i\vec{q}_1\cdot\vec{r}} +
\frac{2J + L}{\sqrt{2}}
\left(\begin{array}{c}
\frac{E-\xi }{\Delta^\ast }\varphi_{\uparrow\, R} \\
-h^{\dagger{\rm T}}_{\downarrow\, L}
\end{array}\right)e^{i\vec{q}_2\cdot\vec{r}} 
\end{eqnarray}
 The excitation $\psi^{9}_{CFL}$ describes the singlet quasiparticle with the
gap $2\Delta $ wheras $\psi^{3}_{CFL},\psi^{8}_{CFL}$ describe octet quasiparticles
with the gap $\Delta $. The probability currents:
\begin{equation}
\vec{j}^A = \psi^{A\dagger }_{particle}\vec{\alpha }\psi^{A}_{particle}+
\psi^{A\dagger }_{hole}\vec{\alpha }\psi^{A}_{hole}
\end{equation}
are conseved and can be used to calculte the transition coefficients
in the scattering procesess.


\begin{thebibliography}{99}
\bibitem{lattice} For example, see Proceedings of {\it Quantum Chromodynamics
and Color Confinement}, edited by H. Suganuma et al. (World Scientific, 2001).
\bibitem{gw_p} D. Gross and F. Wilczek, Phys. Rev. Lett. {\bf 30} (1973) 1343;
H.D. Politzer, Phys. Rev. Lett. {\bf 30} (1973) 1346.
\bibitem{glend} G. Baym and S.A. Chin, Phus. Lett {\bf B62} (1976) 241;
G. Chaplin and M. Nauenberg, Nature {\bf 264} (1976) 235; Phys. Rev. {\bf D16} (1977) 456;
B.D. Keister and L.S. Kisslinger, Phys. Lett {\bf B64} (1976) 117;
W. B. Fechner and P.C. Joss, Nature {\bf 274} (1978) 347.
\bibitem{bailin} F. Barrois, Nucl. Phys. {\bf B129} (1977) 390; 
D. Bailin and A. Love, Nucl. Phys. {bf B190} (1981) 175; Nucl. Phys. {\bf B190} (1981) 751;
Nucl. Phys. {\bf B205} (1982) 119; Phys. Rep. {\bf 107} (1984) 325.
\bibitem{arw1_rssv} M. Alford, K. Rajagopal and F. Wilczek, Phys. Lett. {\bf B422} (1998) 247;
R. Rapp, T. Sch\"afer, E. Shuryak and M. Velkovsky, Phys. Rev. Lett. {\bf 81} (1998) 53.
\bibitem{arw2} M. Alford, K. Rajagopal and F. Wilczek, Nucl. Phys. {\bf B537} (1999) 443.
\bibitem{son} D. Son, Phys. Rev. {\bf D59} (1999) 094019.
\bibitem{bcs} J. Bardeen, L.N. Cooper and J.R. Schrieffer, Phys. Rev. {\bf 106} (1957) 162;
{bf 108} (1957) 1175.
\bibitem{rw} K. Rajagopal and F. Wilczek, Chap. 35 in 
"At the Frontier of Particle Physics/ andbook of QCD", M. Shifman ed. (World Scientific) 
[hep-ph/0011333].
\bibitem{andreev} A.F. Andreev, Zh. Eksp. Teor. Fiz. {\bf 46} (1964) 1823.
\bibitem{sad} M. Sadzikowski, Acta Phys. Pol. {\bf B} (2002)
\bibitem{sad_tac} M. Sadzikowski and M. Tachibana, {\it hep-ph/0201223}, to be published in Phys. Rev. {\bf D}.
\bibitem{carter_reddy} G.W. Carter and S. Reddy, Phys. Rev. {\bf D62} (2000) 103002.
\bibitem{abr} M. Alford, J. Berges and K. Rajagopal, Nucl. Phys. {\bf B558} (1999) 219.
\bibitem{bogoliubov} N. N. Bogoliubov, Zh. Exp. Teor. Fiz {\bf 34} (1958) 58, 73
(JETP {\bf 34} (1958) 41, 51); Y. Nambu, Phys. Rev. {\bf 117} (1960) 648; 
P. G. de Gennes, {\it "Superconductivity of Metals and Alloys"} (Addison-Wesley, New York, 1992).
\bibitem{benistant} P.A.M. Benistant, H. van Kempen and P, Wyder, Phys. Rev. Lett. {\bf 51} (1983) 817.
\bibitem{blonder} G.E. Blonder, M. Tinkham and T.M. Klapwijk, Phys. Rev. {\bf B25} (1982) 4515.
\end{thebibliography}
\end{document}